%% file: main.tex
\documentclass[sigconf]{acmart}

\AtBeginDocument{%
  \providecommand\BibTeX{{%
    \normalfont B\kern-0.5em{\scshape i\kern-0.25em b}\kern-0.8em\TeX}}}

\setcopyright{acmcopyright}
\copyrightyear{2023}
\acmYear{2023}
\setcopyright{rightsretained}
\acmConference[TEI '23]{TEI '23: Proceedings of the Seventeenth
International Conference on Tangible, Embedded, and Embodied
Interaction}{February 26-March 1, 2023}{Warsaw, Poland}
\acmBooktitle{TEI '23: Proceedings of the Seventeenth International
Conference on Tangible, Embedded, and Embodied Interaction (TEI '23),
February 26-March 1, 2023, Warsaw,
Poland}\acmDOI{10.1145/3569009.3573116}
\acmISBN{978-1-4503-9977-7/23/02}

\usepackage{caption}
\usepackage{subcaption}
\usepackage{array, makecell} 
\usepackage{amsmath}
\usepackage{hyperref}
\usepackage{multirow}

\begin{document}

\title{Facebook Data Shield: Increasing Awareness and Control over Data used by Newsfeed-Generating Algorithms}


\author{Jules Sinsel}
\email{j.v.c.sinsel@student.tue.nl}
\affiliation{%
 \institution{Department of Industrial Design, Eindhoven University of Technology}
 \city{Eindhoven}
 \country{The Netherlands}}
 
\author{Anniek Jansen}

\email{a.jansen@tue.nl}
\affiliation{%
 \institution{Department of Industrial Design, Eindhoven University of Technology}
 \city{Eindhoven}
 \country{The Netherlands}}

\author{Sara Colombo}

\email{s.colombo@tue.nl}
\affiliation{%
 \institution{Department of Industrial Design, Eindhoven University of Technology}
 \city{Eindhoven}
 \country{The Netherlands}}


\input{Sections/0_abstract}

\ccsdesc[500]{Human-centered computing~Interactive systems and tools}

\keywords{tangibility, Facebook, social media, user study, data control}


\maketitle
\input{Sections/1_introduction}

\input{Sections/2_relatedwork}

\input{Sections/4_design}

\input{Sections/5_method}
\input{Sections/6_findings}

\input{Sections/7_discussion}

\input{Sections/8_limitations}
\input{Sections/9_conclusion}


\bibliographystyle{ReferenceFormat/ACM-Reference-Format}
\bibliography{references}

\end{document}

%% file: Sections/0_abstract.tex
\begin{abstract}
Social media platforms newsfeeds are generated by AI algorithms, which select and order posts based on user data. However, users are often unaware of what data is collected and employed for this aim, neither can they control it. To open up discussions on what data users are willing to feed the newsfeed algorithm with, we created the Facebook Data Shield, a human-size interactive installation where users can see and control what type of data is collected. By pressing buttons, data categories and/or data variables can be (de)activated. An outer rim with lights gives feedback to users about the level of personalization of the resulting newsfeed. We performed a preliminary study to get insights into what data users are willing to share, their preferred level of control, and the effect of such an installation on users' awareness. Based on our findings, we discuss implications for design and future work.

\end{abstract}

%% file: Sections/1_introduction.tex
\section{Introduction}
Facebook plays an important role in the life of many people, who use it, among others, for social interactions, entertainment, or news consumption. With 2.9 billion users in 2022 \cite{statista2022facebookmostused}, it is the social media platform with the largest user base. While Facebook's services are offered for free to users, revenue is generated by leveraging the data they collect in different ways, including online behavioral advertising (OBA) \cite{Fiesler2018Weraretheproduct}. To do this effectively, Facebook collects and uses data to profile users and serve targeted ads based on such profiles \cite{Habib2022Identifying}. The data they collect usually consist of both “voluntarily” offered information when signing up, a.o. demographic information, and data gathered by monitoring users' online behavior \cite{Joler2016data}. This includes monitoring the behavior on Facebook, e.g., how long one interacts with a post or if they “like” it, but also on other platforms through the use of cookies \cite{Joler2016data}. This process of data collection and processing is usually a black box for users. Even for researchers in related domains, it is not possible to completely map the practices Facebook adopts \cite{Joler2016algorithmes}.

\begin{figure*}
              \centering
      \begin{subfigure}[t]{0.30\textwidth}
         \centering
         \includegraphics[width=\textwidth]{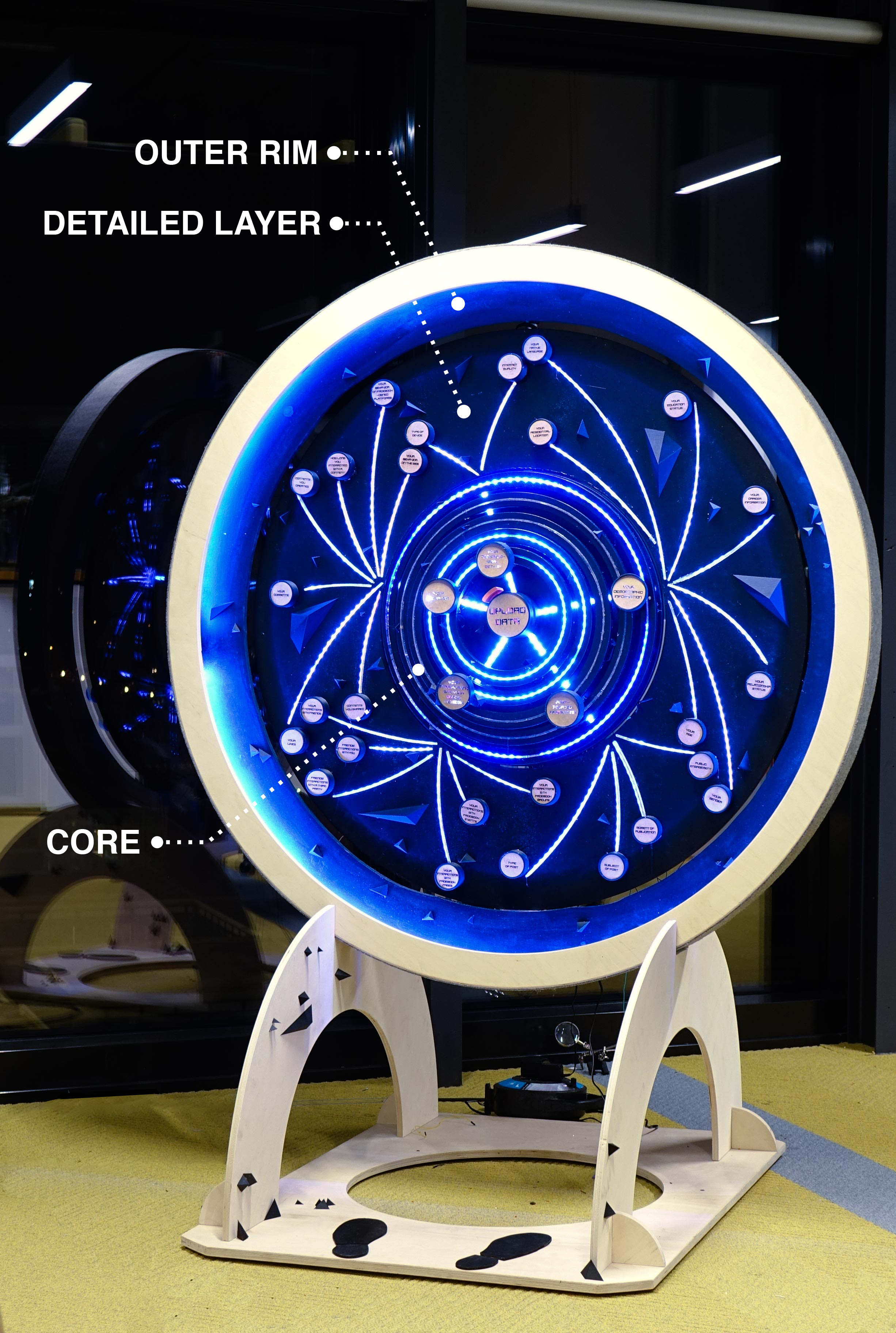}
         \caption{The FDS in its totality, annotated with the different elements}
         \label{fig:installation}
         \Description{Overview photo with the whole FDS visible. Blue lights are visible in the outer rim and between the data variables in the detailed layer and core. There are footprints at the base of the FDS to indicate where people need to stand. There are lines indicating what the outer rim, detailed layer and core is. }
     \end{subfigure}
     \hfill
              \begin{subfigure}[t]{0.30\textwidth}
         \centering
         \includegraphics[width=\textwidth]{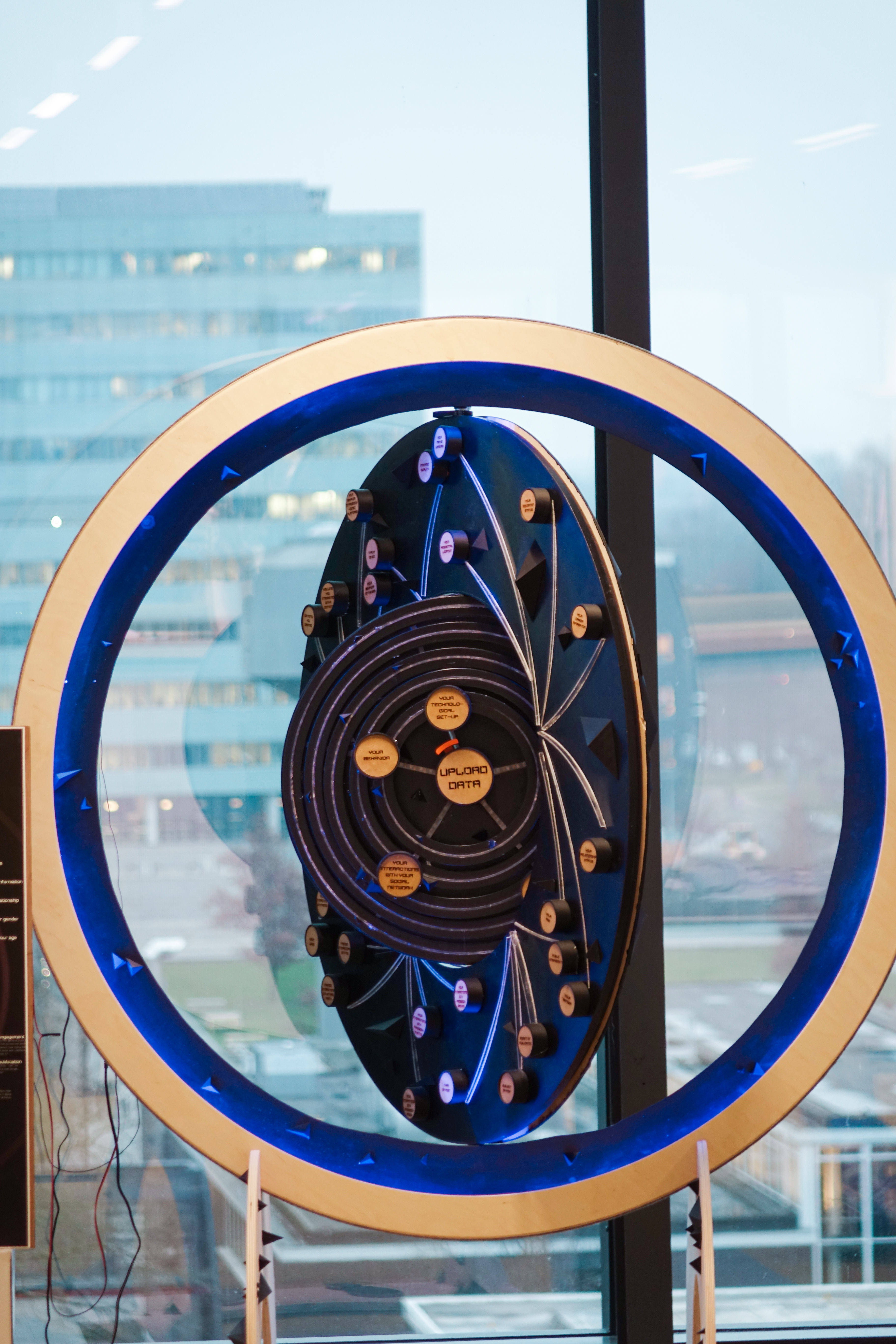}
         \caption{The detailed layer rotating around the core.}
         \label{fig:illustratie}
         \Description{A photo where the detailed layer is partly rotated. The core is fully visible but all lights are off and you see a faint glow of blue in the outer rim.}
     \end{subfigure}
     \hfill
     \begin{subfigure}[t]{0.30\textwidth}
         \centering
         \includegraphics[width=\textwidth]{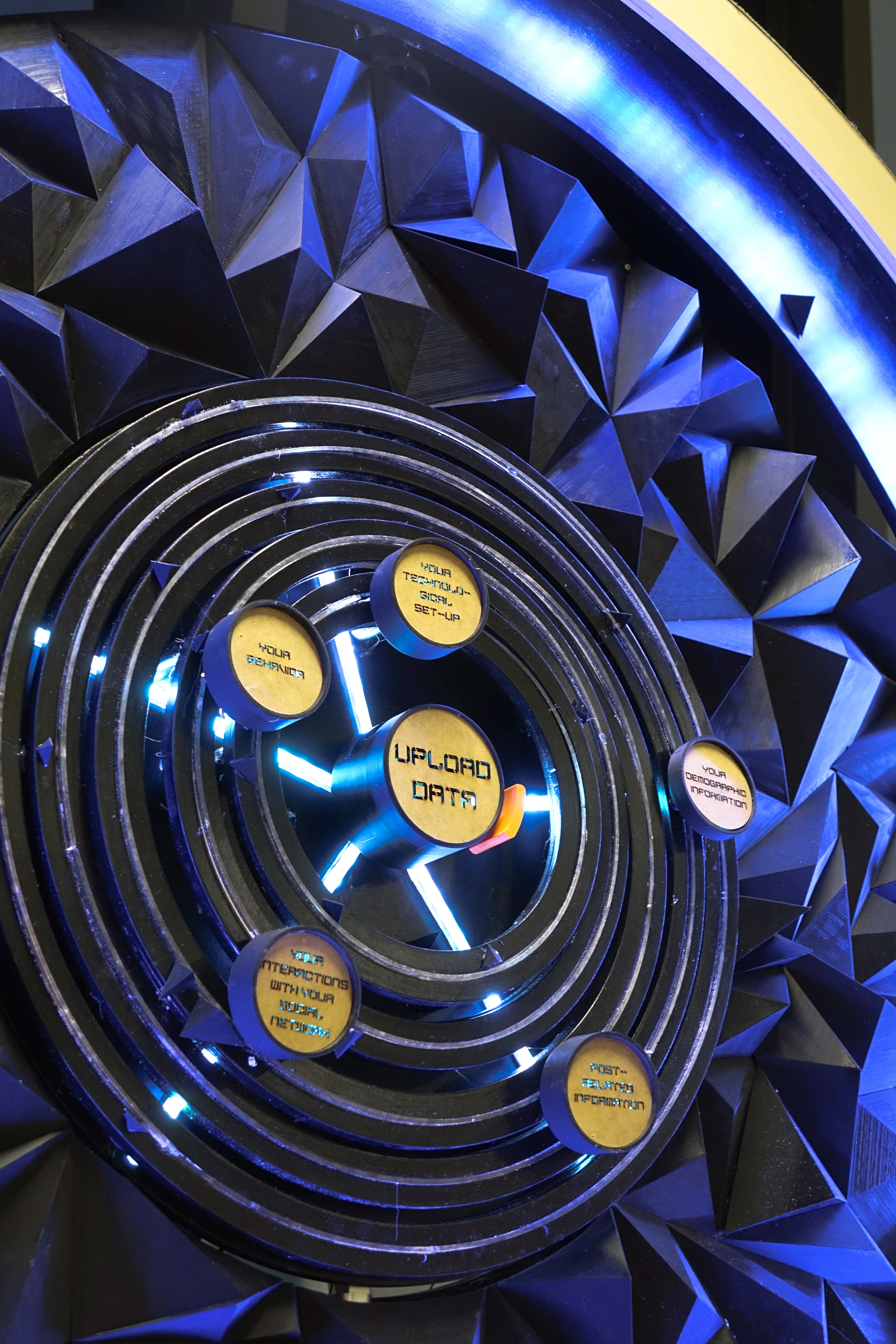}
         \caption{The core of the control panel with the data categories and upload data buttons }
         \label{fig:core}
         \Description{Photo of the five data categories in the core and the upload data button in the middle. Outside the core, you see the black pyramids, which reflect the light from the outer rim. }
     \end{subfigure}
          
     \caption{The Facebook Data Shield}
        \label{fig:FDS}
\end{figure*}

To make OBA even more profitable, major social media platforms, including Facebook, tend to maximize the time users spend on their applications. This is achieved in different ways, including customizing the newsfeed (also called 'feed') to the interests and preferences of users. AI algorithms are employed to predict users' interests and behavior and to filter and order the posts displayed on a newsfeed so that the user feels engaged and compelled to keep interacting with the platform \cite{meta2021how}.

The practices of OBA and newsfeeds personalization raise concerns about data privacy and their effect on society. It has been proven that they can lead to discriminating users \cite{ali2019discrimination}, spreading fake news \cite{Rocha2021impactfakenews, ali2021post} and creating filter bubbles \cite{flaxman2016filter}. Previous studies have shown that people are not always aware of these practices and when they discover them, they often feel uncomfortable with the data that is being collected \cite{pew2019facebook, Habib2022Identifying, Calbalhin2018facebookawareness}. However, a segment of users also considers these targeted ads useful \cite{Habib2022Identifying, Ur2021smartusefulscary}.

Users can exert some control by changing their privacy settings or how their newsfeed is generated. However, these controls are hidden inside nested menus, which makes them not only hard to find but also less known to users \cite{Habib2022Identifying, Habib2020scavengerhunt,Hsu202Awareness}. For example, Facebook currently allows users to choose how their newsfeed is generated, but this is a hidden setting that needs to be activated each time the platform is accessed.

While prior work has identified and analyzed the issues around data control through empirical studies, we adopt a critical design approach \cite{Bardzell2012Criticaldesign} where we aim to disrupt the current practices of data sharing and give full control to users. To achieve this, we designed the Facebook Data Shield (FDS), a human-size physical installation resembling a control panel, which embeds a selection of data points that Facebook currently collects. Users can control what data they want to share with the algorithm that adjusts their newsfeed by interacting with the panel. The FDS consists of three parts (see Figure ~\ref{fig:FDS}): (i) an inner circle, the \emph{core}, showing five general data categories that are being collected (e.g., demographic or behavior information); (ii) a rotating disk around it, the \emph{detailed layer}, which offers the user a more granular control by giving access to specific data variables within each category, and (iii) a lighted outer rim, which visualizes how much control the algorithm has over the user's newsfeed — and the consequent level of personalization of the newsfeed, by varying the light intensity and speed.

To explore what data regular Facebook users want to share and how they use the control panel to select the data they want to share, we ran preliminary studies with 10 industrial design students. Our results show that the layered control panel makes it possible for users to interact with the installation without being overwhelmed with information, while still being provided with enough detail to feel informed and in control. 
The main contribution of this work in progress is the Facebook Data Shield as a design exemplar of the use of critical design and tangible interactions in the context of online data sharing.

%% file: Sections/2_relatedwork.tex
\section{Tangibility and privacy}
Making data or privacy settings more tangible to enhance users' understanding and control of privacy has been used as an approach in previous studies, although the majority of these works focus on privacy in IoT devices \cite{Delgado2022PriKey, Ahmad2020tangibleprivacy, Mehta2018Tangible, Muhander2022PrivacyCube}. For example, PriKey is a small controller in the shape of a key with sliders for control, which offers house inhabitants or visitors control over which sensors are being used in the home, by either disabling them individually or all at once \cite{Delgado2022PriKey}. Another example is the PrivacyCube, which gives users insight into the type of data being collected by active IoT resources, where it is stored, for what purpose, and who has access to the data \cite{Muhander2022PrivacyCube}. 
However, to the best of our knowledge, no work has focused on making the data in social media applications tangible. This is the focus of this work.

%% file: Sections/4_design.tex
\section{Design: Facebook Data Shield}

In this work, we take a different approach from prior research. Instead of performing empirical studies based on the design and testing of real interfaces for privacy control, we adopt a critical design approach \cite{Bardzell2012Criticaldesign} by creating an installation that invites people to engage in a discussion about data control. The installation can be seen as a research probe aimed at both increasing users' awareness and generating knowledge on users' preferences for data control. 

\begin{table*}
    \centering
    \caption{All data categories and variables included in the Facebook Data Shield}
    \label{tab:variables}
    \begin{tabular}{m{15em}|m{20em}  }
    \hline
    Data categories in core & Data variables in detailed layer\\
    \hline
        \multirow{6}{15em}{Your interactions with your social network} & Your interactions with Facebook groups \\ 
        ~ & Your interactions with Facebook events \\ 
        ~ & Your interactions with Facebook pages \\ 
        ~ & Friends' interactions with Facebook pages \\ 
        ~ & Friends' interactions with you \\ 
        ~ & Your interactions with friends \\ \hline
        \multirow{7}{15em}{Your behavior} & Contents you shared \\ 
        ~ & Your likes \\ 
        ~ & Your comments \\ 
        ~ & Contents you created \\ 
        ~ & How long you interacted with a content \\ 
        ~ & Your behavior on the web \\ 
        ~ & Your behavior on Facebook-owned platforms \\ \hline
       \multirow{2} {15em}{Your technological set-up} & Type of device \\ 
        ~ & Internet quality \\ \hline
        \multirow{7}{15em}{Your demographic information} & Your residential location \\ 
        ~ & Your native language \\ 
        ~ & Your education status \\ 
        ~ & Your career information \\ 
        ~ & Your relationship status \\ 
        ~ & Your gender \\ 
        ~ & Your age \\ \hline
        \multirow{4}{15em}{Post-related information} & Public engagement \\ 
        ~ & Moment of publication \\ 
        ~ & Subject of post \\ 
        ~ & Type of post \\ \hline
    \end{tabular}
\end{table*}

\subsection{Design considerations}
Prior work showed that it can be challenging for users to find where they can change their privacy settings \cite{Habib2020scavengerhunt, Habib2022Identifying}. In our design, we eliminated this step, and we only focused on the design of the control itself. By doing so, we intended to explore \emph{'what'} data users are willing to share and at what level of detail they want to control what data is being shared.
The FDS is designed with the vision that it is an object that can be placed in public spaces, where it can attract people to interact with it.

\subsection{Data variables}
To determine which data variables should be included in the FDS, we turned to the work of Joler et al. \cite{Joler2016data}, who created a mapping of the Facebook algorithm, including the data it collects. They identified two clusters of data. The first set of data is the data collected \emph{within Facebook}. This data consists of \emph{Your activities and behavior} on the platform, e.g., \emph{likes, comments, uploaded posts, page visits}, etc., and \emph{your profile information}, this is data you have entered in your account such as your \emph{age, gender, relationship, work}, etc.
The second set of data is \emph{Your digital footprint}, i.e., data collected through your devices and other platforms. Facebook collects this data through cookies and Facebook pixels (small pieces of code that can be embedded in other websites to measure the effectivity of ads and the behavior of visitors) \cite{Metapixel}, but also mobile permissions, i.e., data stored on your phone and information such as your precise location or type of device,  and other companies owned by Meta such as WhatsApp and Instagram \cite{Joler2016data}.

The overview made by Joler et al. \cite{Joler2016data} includes too many data variables (>100) to include in the FDS. Showing users a high number of variables would prevent them from exploring all data points and select the ones they truly want to share. Moreover, it would slow down the interaction, and it may potentially overwhelm the user, decreasing their attention and interest in interacting with the installation. Overall, we decided to favor enabling the user to fully explore and assess a smaller number of data variables, versus giving them a full overview of all data used by the newsfeed algorithm. As identified by Waldsch\"{u}tz and Hornecker \cite{Waldshutz2020}, this process of data curation is an important aspect when representing data physically. To reduce the number of variables, we clustered them (e.g., \textit{number of likes} and \textit{contents of likes} were clustered into a \textit{“likes”} variable) and selected the ones that would give users a sense of the type of data Facebook collects and that could be categorized into overarching data categories.

Since we were interested in the granularity of the control users wish to exercise on their data, we divided the variables into two layers, showing different levels of detail. The first layer includes five general data categories, such as \emph{demographic information} and \emph{technological setup}. The second layer includes more specific data variables belonging to the five categories in the first layer. Examples in this layer for the \emph{Your behavior} category are \emph{Contents you share} and \emph{Your behavior on Facebook-owned platforms}. In total, 26 data variables are included in the detailed layer. A full overview of the selected data variables and to which layer they belong can be found in Table ~\ref{tab:variables}.
On the FDS, the first and second layers are connected by a dynamic light path. When deactivating a variable in the general category layer, the light path and all the corresponding detailed variables are also switched off.

\subsection{Facebook Data Shield}
The FDS consists of three parts (see Figure ~\ref{fig:FDS}): (i) an inner circle, i.e., \emph{the core}, showing the five general categories (see Figure ~\ref{fig:core}), (ii) a rotating disk around it, i.e., \emph{the detailed layer}, displaying the variables from the second layer, and (iii) an outer rim, which gives a light feedback. 

The \emph{core} includes the five data categories as buttons (see Figure ~\ref{fig:core}). Each button is surrounded by a light ring that shows if it is activated (light on) or deactivated (light off). The user can activate or deactivate each variable by pressing the button. 
In the middle of the core is the \emph{upload data} button, which is pressed by the user to mark the end of the interaction and to finalize the variable selection.

The \emph{detailed layer} contains the 26 detailed data variables, represented by 26 buttons on a disk (see Figures ~\ref{fig:illustratie} and ~\ref{fig:installation}). The buttons are connected to the corresponding data category in the core through light strips. Again, each data variable can be (de)activated by pressing the button. As feedback, the light strip will turn on or off. The disk can be rotated 180 degrees, to hide the detailed data variables. The other side shows a landscape of black pyramids designed to reflect light and create a perception of impenetrable obscurity, like the black box algorithm. 

Around the \emph{core} and \emph{detailed layer} is the \emph{outer rim}, which represents how much control the Facebook algorithm has over the user's newsfeed, i.e. the level of personalization (see Figures ~\ref{fig:illustratie} and ~\ref{fig:installation}). In the outer rim, a light moves through the circular profile with different speed and intensity. The more data the algorithm has at its disposal, the brighter and faster the light is.

%% file: Sections/5_method.tex
\section{Method}
To validate our design and get an initial understanding of how it could be used to trigger discussions about the Facebook algorithm, we conducted preliminary studies. 

\subsection{Participants}
Our goal is to deploy the FDS in a public space where a variety of participants can interact. However, for this preliminary evaluation, we recruited participants through convenience sampling among industrial design students from the same university, with the inclusion criteria of being weekly users of Facebook. This ensured that participants would have an understanding of the Facebook newsfeed and could reflect on their own practices regarding data sharing. In total 10 participants (male (n=5), female(n=5), age: M=23.6, SD=1.07) took part in the study.

\subsection{Materials and setup}
The FDS was used during all sessions, and the final settings were recorded manually. A poster was placed beside the FDS with short descriptions for each variable. The FDS was set up in a large open space. The interviews were conducted next to the installation so that participants could look at it. 

\subsection{Procedure}
The study consisted of three parts: an introduction, the interaction with the installation, and a semi-structured interview. The study protocol complied with the University Ethical Review Board procedures and all data were managed in accordance with GDPR regulations.

The sessions started with the participants giving consent. They were asked their age, gender, educational background and for what purpose they used Facebook. Next, they received a brief introduction. Participants were explained that the Facebook newsfeed is generated by an algorithm that uses their data to determine what posts will be most relevant to them. The introduction concluded with describing the installation scope and interaction modalities.

After the introduction, participants were invited to change the settings in the FDS according to their data sharing preferences. At the start, the \emph{detailed layer} was hidden so that only the five data categories in the \emph{core} were visible. Participants could rotate the \emph{detailed layer} whenever they wished, or could decide not to do so. The session ended when the participant pressed the \emph{upload data} button.
During the interaction, they were asked to think aloud, and a researcher observed the interactions and took notes.

The session ended with a semi-structured interview. Three questions were asked, to investigate: i) how participants felt when they were given the power to determine what data Facebook collects/uses, ii) which variable triggered them the most to deactivate and why, iii) whether and how often they would use a similar feature if it were available on Facebook.

\subsection{Data analysis}
All sessions were audio recorded and transcribed. The interactions with the buttons were recorded as active-passive describing whether the participant interacted with the variable and activated-deactivated to notate the final setting of each variable. The quantitative interaction data were aggregated and analyzed. 
The qualitative data of the recordings and observation notes was analyzed using a thematic analysis \cite{braun2012thematic}. 

%% file: Sections/6_findings.tex
\section{Findings}

\begin{table*}
    \centering
   
    \caption{An overview of which data categories were kept activated (1) and were deactivated (0) for each participant }
     \label{tab:uploadsettings}

    \begin{tabular}{l|l|l|l|l|l|l|l|l|l|l|c}
    \hline
        ~ & P1 & P2 & P3 & P4 & P5 & P6 & P7 & P8 & P9 & P10 & Total active \\ \hline
        Your interactions with your social network & 1 & 1 & 1 & 1 & 1 & 0 & 0 & 1 & 0 & 1 & 7/10\\ 
        Your behavior & 0 & 1 & 1 & 0 & 0 & 0 & 0 & 0 & 0 & 0 & 2/10 \\ 
        Your technological set-up & 1 & 1 & 0 & 0 & 0 & 1 & 1 & 0 & 0 & 1 & 5/10\\ 
        Your demographic information & 0 & 0 & 1 & 1 & 0 & 1 & 1 & 1 & 0 & 1 & 6/10 \\ 
        Post-related information & 0 & 1 & 1 & 0 & 1 & 1 & 1 & 0 & 0 & 1 & 6/10\\ \hline
    \end{tabular}
\end{table*}

\subsection{Control level}
During the study, participants decided themselves if and when to turn the outer circle to access the \emph{detailed layer}. The majority of the participants (8/10) looked at the \emph{detailed layer}, but only seven participants interacted with it (in total 24 changes were made to the detailed layer, versus 24 to the core). P1, P3 and P10 only used the core although P3 did look at the detailed layer. However, they explained during the interaction that the basic layer provided them with enough control. The rest of the participants did use the detailed layer to control their data in more detail and to receive more information about what each data category entailed. P6  felt that the \emph{detailed layer} was needed because the core could hide too much information. 

The number of changes made in the detailed layer differed between participants and data categories. For the data category \emph{Your interactions with your social network} most changes were made in the detailed layer (13 out of the 60 possible changes for that data category in the detailed layer were changed) while for \emph{Your behavior} only three changes were made (out of 70 possible changes). Overall, when interacting with the detailed layer, participants tended to stick to the settings resulting from their choices in the main data category (i.e., all variables either activated or deactivated). However, seeing the specific data variables triggered new reflections and resulted occasionally in (de)activating some of them. For example, P2 deactivated \emph{Your interactions with Facebook events} and \emph{Friend's interactions with a third party} since they found them \emph{“annoying”}. 

While most participants accessed the detailed layer, they also expressed that they would expect to be overloaded with information if there was no separation between the two layers or if the detailed layer would be visible from the start. Even by splitting the two layers, several participants had an overload of information when presented with the second detailed layer.

\subsection{Motivators: privacy, newsfeed personalization, and goals}
When making decisions about which variables to turn on or off, different motivations were used.

Firstly, participants claimed that the reason for deactivating certain data categories and/or variables was to protect their privacy. 
\emph{Your behavior} was considered by most participants (6/10) the most personal data, and therefore they wished to deactivate them (see Table ~\ref{tab:uploadsettings}). They felt that this type of data was an invasion of their privacy. As P5 and P7 expressed, \emph{“Facebook has no right to know this information”} (P5) and  \emph{“Creepy, you don't know how they use it”} (P7). Demographic information was also seen as private, but it was not deactivated as often (40\%).

A second motivation for (not) changing settings was the influence of data sharing on the level of personalization and consequent the quality of the newsfeed. This reflection mainly resulted in not deactivating some data. While one participant deactivated all data, the others tried to find a balance between deactivating data (for privacy reasons) but still having a relevant and engaging newsfeed. For example, P3 left \emph{Your behavior} and \emph{Your demographic information} active since they expected them to be necessary for a personalized newsfeed  showing relevant posts. 

These choices were often linked to the participant's goal in using Facebook. Five participants used it for entertainment, four for social interactions and one for business. Those using it for entertainment were less likely to deactivate variables (they deactivated between 2-17 detailed variables each). Social users deactivated between 15-19 detailed variables each, and the business user deactivated all variables.

\subsection{Control and engagement}
The FDS was designed to both offer users a sense of control over their data and engage them in a critical reflection. Participants were unanimous in that the FDS provided them with control, and they all indicated they would use this option if it were to be implemented on Facebook. Moreover, they also indicated that it was \emph{“eyeopening”} (P5) and encouraged self-reflection (P7).

%% file: Sections/7_discussion.tex
\section{Discussion}
Based on our preliminary findings, we reflect upon the design of the FDS and discuss implications for design.

With the FDS, we aimed to explore how the hidden intangible data settings of Facebook could be made visible and tangible. We opted for a design where data variables are represented as buttons that can be (de)activated and are divided into two layers of detail. Our findings showed that using different levels of details has potential, since it first introduces users to the general types of data being collected without overwhelming them, and then provides a more granular control over their settings through the \emph{detailed layer}. At the same time, our results also show that only a few changes were made to the variables in the detailed layer. A potential explanation for this is that after setting the data group, the control in the complex layer becomes superfluous, since their feeling to each of these variables is the same as their feeling over the entire data group. However, having made a preliminary choice might also affect users' willingness to revise that decision later on, or to reflect on the differences in the detailed variables. This might suggest that data categories with more variability in the detailed layer, e.g., \emph{your interactions with your social network}, should be split into additional main data categories that represent them better, if the control happens at the core level.
However, reducing the control to only the data categories might leave participants feeling uninformed about the specific contents of each category, as participants appreciated the detailed variables as providing additional explanation. An alternative would be to give users the option to uncover one data category at the time, instead of fully rotating the \emph{detailed layer}. To facilitate this, the rotating mechanism of the FDS should be redesigned.

Presenting the data as tangible and actionable resulted in participants feeling in control, which was highly appreciated. This confirms earlier findings about people wanting to be in control of their data \cite{Hsu202Awareness, Habib2022Identifying}.

While it seems unfeasible to offer social media users tangible control over their data, this approach proved to be useful to discuss certain aspects of data collection and control, and to understand users' needs, without focusing on where such a control can be found. Findings of this study can help to better understand users' wishes around privacy and the desired level of data awareness and control. For instance, the use of basic and detailed data layers to provide nuanced control and to inform users seems to be appreciated, and it could be easily translated into the digital world. While Facebook currently offers limited control, much more can be done to empower the user and give them agency in the use of their data.

Future studies might adopt a similar approach to investigate users' data control on other social media platforms, as the general design of the FDS supports multiple platforms. Only the data categories and variables are Facebook specific.

%% file: Sections/8_limitations.tex
\section{Limitations and Future Work}
One of the main limitations of this work is the small and homogenous study sample. A larger and more diverse sample is needed to find relevant patterns in the interaction data. A future study is already planned where the installation will be deployed at a large public event. This will result in a larger and more diverse set of participants and a greater amount of data being collected.

Finally, the FDS represents the level of personalization of the resulting newsfeed - and the consequent expected level of user's engagement,  in an abstract manner through light, since it is not possible to adjust the actual Facebook newsfeed of users. This resulted in users having to imagine the quality of their newsfeed, instead of experiencing it. Depending on the user's familiarity with and knowledge of Facebook, their understanding of the effects of their data selection can be different. Nevertheless, we argue that our design offers insight in what data users are willing to share and the desired level of control. Moreover, the FDS increases users' awareness, and it fosters reflection on the delicate balance between privacy and personalization or engagement.

%% file: Sections/9_conclusion.tex
\section{Conclusion}
Facebook collects a wide range of data from users. However, what data exactly is collected and how this can be controlled is often not clear. In this work, we present the Facebook Data Shield, a human-size tangible installation aimed at inviting users to start a discussion and reflection about data sharing and data control. 
We ran a preliminary study with 10 Industrial Design students who are frequent Facebook users. Our findings show that such a tangible interactive installation can increase users' awareness, stimulate reflection, and provide insight into what participants consider too personal to share and how they prefer to control their data. Results also hints to a possible correlation between users' goal and their choices in data sharing. We will further investigate the potential of this installation in exploring these aspects in a future study with a larger and more diverse group of participants, to build on these preliminary findings. 